\documentclass[11pt,letterpaper]{JHEP3}
\usepackage{epsfig}
\def\fin{{f_\infty}}
\def\lp{\ell_P}
\newcommand{\X}{\mathcal{X}}
\newcommand{\Z}{\mathcal{Z}}

\newcommand{\be}{\begin{equation}}
\newcommand{\ee}{\end{equation}}
\newcommand{\bea}{\begin{eqnarray}}
\newcommand{\eea}{\end{eqnarray}}
\newcommand{\ba}{\begin{eqnarray}}
\newcommand{\ea}{\end{eqnarray}}

\newcommand{\beq}{\begin{equation}}
\newcommand{\eeq}{\end{equation}}
\newcommand{\beqa}{\begin{eqnarray}}
\newcommand{\eeqa}{\end{eqnarray}}
\newcommand{\beqar}{\begin{eqnarray*}}
\newcommand{\eeqar}{\end{eqnarray*}}

\newcommand{\la}{\lambda}

\title{On higher derivative gravity, $c$-theorems and cosmology}
\author{Aninda Sinha \\
\it Perimeter Institute for Theoretical Physics\\
\it Waterloo, Ontario N2L 2Y5, Canada\\
\vskip .5cm

{\rm E-mail:}\ \ {\tt asinha@perimeterinstitute.ca}}

\abstract{We consider higher derivative gravity lagrangians in 3 and 4 dimensions, which admit simple $c$-theorems, including upto six derivative curvature invariants. Following a suggestion by Myers, these lagrangians are restricted such that the fluctuations around (anti) de Sitter spaces have second order linearized equations of motion. We study $c$-theorems both in the context of AdS/CFT and cosmology. In the context of cosmology, the monotonic function is the entropy defined on the apparent horizon through Wald's formula. Exact black hole solutions which are asymptotically (anti) de Sitter are presented. An interesting lower bound for entropy is found in de Sitter space. Some aspects of cosmology in both $D=3$ and $D=4$ are discussed. }


\begin{document}


\tableofcontents
\section{Introduction}
 It is usually difficult to deal with equations of motion with more than two derivatives. Typically such theories are plagued with ghosts \cite{ghosts,Stelle:1977ry} . String theory has in principle, a systematic way to compute higher derivative corrections to the low energy effective action. For instance it is well known that heterotic string has a correction that is quadratic in curvature \cite{gwitt,zwie}. If one computed fluctuations with this effective action, one would find ghosts. A conservative viewpoint to this problem is to argue that field redefinitions allow us to write the quadratic correction as a Gauss-Bonnet term in which case this problem disappears \cite{zwie}. It is difficult to imagine that something similar can be done order by order around an arbitrary background. Of course one could take the attitude that there are an infinite set of such terms and once all of them are taken into account the mass of the problematic ghost modes would be pushed to infinity or essentially they would be removed from the spectrum. This would only work if the higher derivative terms are treated perturbatively.

 In this paper, we will consider higher derivative lagrangians in $D=3$ and $D=4$  that include upto six derivative terms, i.e., the action schematically reads
 \be
 I\propto \int d^D x\, \sqrt{-g}(R-2\Lambda+ \tilde\lambda R^2+ \tilde\mu R^3)\,,
 \ee
 where $R^2$ and $R^3$ denote a general set of four and six derivative curvature corrections. The effect of stringy $R^2$ corrections were first studied in \cite{bd}. While $R^2$ corrections arise in supersymmetric string theory, $R^3$ corrections arise in nonsupersymmetric string theories \cite{nonsusyr3}. Six derivative theories have featured recently in \cite{mr,me,Oliva:2010zd,mps,quasitop}. Following a suggestion by Myers \cite{mr, future1}, one of the main goals of this paper is to construct general lagrangians such that when one considers fluctuations around (anti) de Sitter space, the linearized equations of motion are two derivative. We will not treat the higher derivative terms perturbatively in $\tilde\lambda$ or $\tilde\mu$. One of the main reasons for not treating the higher derivative corrections perturbatively is to leave the possibility of probing issues such as lowering the viscosity bound \cite{shenker,mps} in consistent toy models open. We will also demand that the resulting lagrangians admit for simple $c$-theorems in AdS/CFT \cite{future1} and cosmology. If there is a flow between two theories, then the value of the $c$-function at the fixed points is supposed to be a measure for the number of degrees of freedom at the fixed points.

 The degrees of freedom on the CFT side are captured by central charges. In 1+1 dimensional CFTs, unitarity, a conserved stress energy tensor and the Euclidean group of symmetries are enough to show the existence of a $c$-theorem whereby the ultraviolet value of the central charge is greater than the infrared value \cite{zam}. Using the AdS/CFT correspondence, it is straightforward to show this $c$-theorem using gravity equations of motion and the null energy condition \cite{cthm}. In \cite{me}, postulating the existence of a simple $c$-theorem was used to derive the new massive gravity model \cite{tmg, nmg,nmg2,nmg3,mgads,mgads2,mgads3}. These theorems were considered further in \cite{cthms3D}. In \cite{ms}, $c$-theorems in arbitrary dimensions were investigated and an entanglement entropy interpretation for the quantity that was flowing was given in arbitrary dimensions. Related issues have been further discussed in \cite{follow}. In the derivation of the theorem, the equations of motion and null energy conditions are used--the cosmological constant does not enter in an important way. It is natural to ask if a similar theorem exists in the context of cosmology.

 A $c$-theorem for cosmology similar in spirit \cite{cthm} to the AdS/CFT correspondence was first proposed by Strominger in \cite{andyds}. Here the starting point is to assume that the bulk metric is given by
 \be
 ds^2=-dt^2+ a(t)^2 d{\bf x}_{D-1}^2\,,
 \ee
 such that the scale factor at early and late times behaves like
 $\frac{\dot a}{a}\rightarrow H_i , \frac{\dot a}{a}\rightarrow H_f$ respectively,
with $H_i$ being the inflationary era Hubble constant and $H_f$ being the current value of the Hubble constant. In the intermediate stages, the evolution is governed by standard FRW equations.  In two derivative Einstein gravity, the FRW equations give us
\be\label{2der}
\dot H=\frac{\ddot a}{a}-(\frac{\dot a}{a})^2=-8\pi G(\rho+P)\,,
\ee
so that if the null energy condition $\rho+P\geq 0$ holds  then
\be\label{sdef}
s(t)\sim \frac{1}{H(t)^{D-2}}
\ee
will be an increasing function in time\footnote{In the context of de Sitter/CFT correspondence, late times in the bulk correspond to ultraviolet in the supposed boundary field theory while early times correspond to the infrared. Bulk time evolution is thought to be an inverse RG flow (IR to UV) and hence corresponds to ``integrating in" degrees of freedom. Hence it is natural to expect that entropy increases with time.}. Here $s(t)$ is interpreted as the entropy on the apparent horizon. We will be interested mostly in the flat FRW case where the apparent horizon is the Hubble horizon. When the space is de Sitter, the notion of cosmological horizon and apparent horizon coincide. Thermodynamics aspects of apparent horizons have been studied in \cite{akbarcai} and extended to Gauss-Bonnet and Lovelock theories in higher dimensions. In the context of dS/CFT, $c$-theorems have been investigated in \cite{morecds}. Of course as it stands eq.(\ref{2der}) does not prevent $a(t)$ from running off to zero in the early past. It is natural to expect that quantum corrections will be very important in this case and there will be corrections to eq.(\ref{sdef}).

A first step towards understanding quantum effects is to study the inclusion of higher derivative curvature corrections in the action. Typically this may lead to problems for instance with ghosts although it is possible that the approach of \cite{ghosts} in dealing with ghosts will still allow us to extract useful physical information. In 2+1 dimensions, an interesting higher derivative gravity theory was proposed in \cite{nmg,unit, no} where four derivative terms $R_{ab}R^{ab}-3/8 R^2$ are added to the Einstein-Hilbert lagrangian. The propagating degree of freedom is a massive spin-2 field. Around flat space, this degree of freedom can be shown to be unitary in spite of the appearance that the equations of motion are intrinsically higher order. When the equations of motion are higher order, schematically the graviton propagator looks like $1/(p^2(p^2+m^2))\propto 1/p^2-1/(p^2+m^2)$ so that there is always an additional degree of freedom which the wrong sign kinetic term. In 2+1 dimensions it is possible to make the non-propagating mode have the wrong sign while the propagating massive mode have the right sign \cite{unit}. While this seems to work around flat space, this construction is problematic in the context of AdS/CFT. In this case, demanding that the massive mode is unitary in the bulk leads to the dual CFT have negative central charge \cite{mgads}!

One possible way out of this is to add more terms to the lagrangian as in \cite{me} and tune the coefficients such that the equations of motion for the fluctuations around AdS space is two derivative \cite{paulos}. Then for a specific choice of the parameters it is possible to show that the CFT central charge is positive while the bulk theory is unitary. Of course this works only around (anti) de Sitter space and it is not clear how severe the problem with unitarity will be around other spacetimes. In any event, it is fair to say that this construction is an interesting one and worth probing further. The way that one gets two derivative equations of motion in this case is that the six derivative $R^3$ terms cancel off the offending higher derivative terms arising from the four derivative terms. This of course only works around a non-trivial background and a similar construction {\it cannot} be used around flat space.

We want to construct an interesting class of higher derivative models in $D=4$ such that fluctuations around (anti) de Sitter space have two derivative equations of motion. There are several motivations behind doing this.
\begin{enumerate}
\item We wish to propose interesting higher derivative lagrangians in $D=4$. Had we worked just with four derivative $R^2$ lagrangians we would be led to the Gauss-Bonnet term which does not alter the equations of motion in $D=4$ as it is a total derivative. We wish to do something more interesting than this. In particular, we want to allow for the possibility of exact black hole solutions as in \cite{mr}. However, the approach used in \cite{mr} does not extend to $D=4$.
\item Higher derivative gravity is an interesting playground to consider transport properties in interesting field theories at strong coupling . It has been used to study bounds on the ratio of shear viscosity to entropy density in the context of quark gluon plasma \cite{rev,mps}. An unsolved question here is if the ratio can be driven to zero without any pathologies on the gravity side. No useful study has been yet carried out in the context of $d=2+1$ CFTs which are useful in the AdS/CMT applications. Part of the reason is that no straightforward generalization of Gauss-Bonnet or quasitopological gravity exists in $D=4$.
\item We wish to consider $c$-theorems in this context both in AdS/CFT and cosmology. These were studied in higher derivative gravity in \cite{ms,future1} in the context of AdS/CFT.
\item $f(R)$ theories have been extensively studied as viable alternatives to inflation \cite{fofR, no1}. Generalizations to $f(GB)$ or as functions of Gauss-Bonnet have also been considered \cite{fofG}. The $f(GB)$ models may have some problems since they appear to be incompatible with observations \cite{fofR}. Thus (and otherwise) it is interesting to look for alternatives.
\end{enumerate}

Keeping these motivations in mind we will engineer higher derivative lagrangians including upto six derivative curvature invariants in $D=4$. These can be thought to be distant cousins of the quasitopological theory of Myers and Robinson \cite{mr}. We will find a five parameter family of such lagrangians which yield two derivative equations of motion for fluctuations around (anti) de Sitter spaces and which allow a simple $c$-theorem as in \cite{me,ms}. A two parameter subspace is found such that the equations of motion for fluctuations around FRW or a static domain wall are two derivative. This lagrangian coincides with the choice of Gauss-Bonnet for the $R^2$ terms and a cubic invariant of the Weyl tensor for the $R^3$ terms. If one considered black holes in such spacetimes, the solutions would receive corrections from the six derivative terms. A three parameter subspace is found where exact black hole solutions exist. Rather interestingly, we will find that there is a lower bound for the entropy in de Sitter spaces. Some aspects of this theory will also be discussed in the context of AdS/CFT $c$-theorems in \cite{future1}.

This paper is organized as follows. In section 2, we write down the equations of motion to be used in the rest of the paper and specify our conventions. In section 3, we review the construction of new massive gravity and its extensions. We consider $c$-theorems in the context of cosmology. In section 4, we turn to $D=4$. After constructing the lagrangian, some exact black hole solutions are presented. In section 5, we turn to discussing $c$-theorems in the context of cosmology. We conclude with a discussion of open problems in section 6.

 \section{Six derivative theories and equations of motion}
 We will be interested in lagrangians that include upto six derivative curvature invariants. These take the form \cite{mr}
 \be
I=\frac {1}{2\lp^{D-2}}\int d^{D}x \sqrt{-g} \left[\pm\frac{(D-1)(D-2)}{L^2}+R+ L^2 \X_4+ L^4 \Z_4\right] \label{action}
\ee
where
 \beqa
 \X_4&=&\lambda_1 R_{abcd}R^{abcd}+\la_2 R_{ab}R^{ab}+\la_3 R^2\ ,
 \label{euler4}\\
 \Z_4&=&
\mu_1 R_{abc}{}^d R^b{}_{edf} R^{aecf} +
\mu_2 R_{a b c d}R^{a b c d} R+\mu_3 R_{a b c d}R^{a
b c}{}_{e}R^{d e}
  \label{result5a}\\
&&\qquad\mu_4 R_{a b c d} R^{a c}R^{b d}
+\mu_5 R_a{}^{b}R_b{}^{c}R_c{}^{a} +\mu_6 R_a^{\,\,b}R_b^{\,\,a}R
+\mu_7 R^3\ .\nonumber
 \eeqa
We have left out terms that involve $\nabla R$'s. There are several reasons for this. Firstly, as shown in \cite{mr}, in order to get exact black hole solutions, these terms should be absent. Secondly, had these terms been present, schematically they would look like $(\nabla R)(\nabla R)$ so that when we vary this, there would be contributions that look like $R \nabla \nabla \delta R$ so that the equations of motion for fluctuations would necessarily involve more than two derivatives unless these contributions canceled among themselves. We will assume in what follows that there are no such terms although including them should not be a problem so long as their contributions to fluctuations cancel among themselves.

The equations of motion that follow from this are given by \cite{mr,Oliva:2010zd}
\be
R_{ab}-\frac{1}{2}g_{ab}R\mp \frac{(D-1)(D-2)}{2L^2}g_{ab}-L^2 K^{(2)}_{ab}-L^4 K^{(3)}_{ab}=0\,,
\ee
where
\begin{eqnarray}
K^{(2)}_{ab}&=& \lambda_3(-2 R R_{ab}+2 \nabla_a \nabla_b R+ g_{ab}[\frac{1}{2} R^2-2 \nabla^2 R])\nonumber \\&&+\lambda_2 (-2 R_a^c R_{cb}+2 \nabla_c \nabla_{(a} R_{b)}^c-\nabla^2 R_{ab}+g_{ab}[\frac{1}{2} R_{cd}R^{cd}-\frac{1}{2}\nabla^2 R])\nonumber \\
 &&+
  \lambda_1 (\frac{1}{2}g_{ab}R_{cdef}R^{cdef}-2 R_{a cde}R_{b}^{\ cde}-4\nabla^2 R_{ab}+2 \nabla_{a}\nabla_{b}R+4 R^c_{\ a}R_{b c}+4 R^{c d}R_{c (ab) d})\,,\nonumber \\
 \eea
while
\beqa
&&\!\!\!\!\!\!\!\!\!\!K^{(3)}_{ab}\nonumber \\ &=& \mu_1 (-3 R_{dec}^{\ \ \ \ f} R^{cge}_{\ \ \ a} R_{fgb}^{\ \ \ d}+3 \nabla_d \nabla_c R_{e \ f (b}^{\ d}R_{a)}^{\ \ e c f}-3\nabla_c \nabla_d R_{(a \ b)}^{\ \ e \ \ f}R_{e \ f}^{\ d \ c}+\frac{1}{2}g_{ab} R_{c \ e}^{\ d \ f}R_{d \ f}^{\ g \ h}R_{g \ h}^{\ c \ e})\nonumber \\
&& +\mu_2 (-2 R R_{acde} R_b^{\ c d e}-R_{ab}R_{cdef}R^{cdef}+\nabla_b\nabla_a R_{cdef}R^{cdef}+4 \nabla_d\nabla_c R R_{a \ \ b}^{\ cd}\nonumber \\ &&~~~~~~~~~+\frac{1}{2}g_{ab}[R R_{cdef}R^{cdef}-2\nabla^2 R_{cdef}R^{cdef}])\nonumber\\
&&+\mu_3 (-2 R_{acd}^{\ \ \ e}R_b^{\ cdf}R_{ef}-R^{defc}R_{def(a}R_{b)c}-\nabla_f\nabla_{(b}R_{a)}^{\ \ e c d} R_{cde}^{\ \ \ f}-\frac{1}{2}\nabla^2 R^{cde}_{\ \ \ a}R_{cdeb} \nonumber \\ &&~~~~ +2\nabla_d \nabla_c R^{edc}_{\ \ \ (a}R_{b)e}+2 \nabla_d\nabla_c R^{e\ \ \ \ c}_{\ (ab)}R_e^{\ d}
+\frac{1}{2}g_{ab}[R^{cdef}R_{cde}^{\ \ \ g}R_{fg}-\nabla_f\nabla_c R^{degc}R_{deg}^{\ \ \ f}])\nonumber \\
&&+ \mu_4 (-3 R_{c(a}R^{\ \ \ c}_{b)d \ e}R^{de}+2 \nabla_d \nabla_{(b} R_{a) c \ e}^{\ \ \ d}R^{ce}-\nabla^2 R_{acbd}R^{cd}+\nabla_d\nabla_c R_{(a}^{\ \ c}R_{b)}^{\ \ d}-\nabla_d\nabla_c R^{cd}R_{ab} \nonumber \\
&&~~~~~~~~~+\frac{1}{2}g_{ab}[R^{cd}R_{cedf}R^{ef}-\nabla_d\nabla_c R_{e \ f}^{\ c \ d}R^{ef}])\nonumber \\
&&+\mu_5(-3 R_{ac}R^c_d R^d_b-\frac{3}{2} \nabla^2 R_{ac}R^c_b +3 \nabla_c\nabla_{(a} R_{b)}^d R_d^c+g_{ab}[\frac{1}{2}R_c^d R_d^e R_e^c-\frac{3}{2} \nabla_d \nabla_c R^c_e R^{ed}])\nonumber \\
    &&+\mu_6(-R_c^d R_d^c R_{ab}-2 R R_{ac}R^c_b +\nabla_b \nabla_a R_{cd}R^{cd}-\nabla^2 R R_{ab}+2 \nabla_c \nabla_{(b}R_{a)}^c R\nonumber \\ && ~~~~~~~~~~+g_{ab}[\frac{1}{2} R R_{cd}R^{cd}-\nabla_d\nabla_c R R^{cd}-\nabla^2 R_{cd}R^{cd}])\nonumber \\
    &&+\mu_7 (-3 R^2 R_{ab}+3 \nabla_a \nabla_b R^2+g_{ab}[\frac{1}{2}R^3-3 \nabla^2 R^2])\,.
\eeqa
For computational purposes, it is sometimes easier to use the effective action approach as described in \cite{mr}. Suppose one is interested in finding planar black hole solutions. Here one starts with an ansatz for the metric of the form
\be
ds^2=-N(r)^2 f(r)dt^2+\frac{dr^2}{f(r)} + \frac{r^2}{L^2} (dx^2+dy^2)\,,
\ee
plugs this into the action and works out the equations of motion for $N(r)$ and $f(r)$. When we deal with fluctuations around a given background, we turn on all metric components for the fluctuations with arbitrary spatial and temporal dependence, expand the action upto second order and work out the equations of motion for the fluctuations.

Finally we make a note of the following relations with the invariants of the Weyl tensor $C_{abcd}$ in $D=4$. At quadratic order, there is a unique invariant which is related in the following way:
\be
C_{abcd}C^{abcd}=R_{abcd}R^{abcd}-2 R_{ab}R^{ab}+\frac{1}{3} R^2\,,
\ee
At cubic order there are two invariants
\be\label{Winv}
W_1= C^{rstu} C^{vw}_{ \ \ \ r t}C_{svuw}\,,\quad W_2= C^{rstu} C_{\ r \ \ t}^{v \ w}C_{swuv}\,.
\ee
$\alpha W_1+\beta W_2$ can be expanded in terms of $\Z_4$ by choosing the $\mu_i$'s as follows:
\be
\mu_1=\frac{\alpha+\beta}{2}\,,\quad \mu_2=\frac{\alpha-5\beta}{16}\,,\quad \mu_4=-{3\beta}=-2 \mu_3\,,\quad \mu_5=8\mu_2=-\mu_6\,,\quad \mu_7=\frac{11\alpha-43\beta}{144}\,.
\ee
Now there is a Schouten identity that leads to the fact that $W_1=W_2$ in $D\leq 5$. As a result, in order to be consistent, it had better be true that
\be\label{idX}
{\cal X}_5=R_{a b c d}R^{a b c d} R-4 R_{a b c d}R^{a
b c}{}_{e}R^{d e}+8 R_{a b c d} R^{a c}R^{b d}
+8 R_a{}^{b}R_b{}^{c}R_c{}^{a} -8 R_a^{\,\,b}R_b^{\,\,a}R
+ R^3=0\,.
\ee
Where does this identity come from? This comes from the fact that in 5 dimensions we can construct
\be
\epsilon^{abcde}\epsilon^{fghij}R_{abfg}R_{cdhi}R_{ej}=-4 {\cal X}_5\,,
\ee
and this should vanish in four dimensions. Thus using this identity we can drop one of the $\mu_i$'s for $i\geq 2$. For example, we could choose to drop $\mu_7 R^3$. In what follows we will set $\beta=0$ when $D=4$ but will retain all the $\mu_i$'s. This will serve as a cross-check on the algebra.

\section{Extensions of new massive gravity}
In this section we review the construction of six derivative lagrangians in \cite{me} in $D=3$. In $D=3$, the Riemann tensor is given in terms of the Ricci scalar and Ricci tensor. As a result the number of independent cubic terms reduces to just three. These lagrangians were constructed so that a simple $c$-theorem existed on using the null-energy condition. The action is given by
\be \label{actmain}
I=\frac{1}{2\lp}\int d^3 x \sqrt{-g}(R+\frac{2}{L^2}+L^2 {\mathcal R}_2+L^4 {\mathcal R}_3)\equiv \frac{1}{2\lp}\int d^3 x \sqrt{-g}(R+\frac{2}{L^2}+K) \,,
\ee
where
\begin{eqnarray}\label{lagmain}
{\mathcal R}_2&=& 4(\lambda_1 R_{a b}R^{a b}+\lambda_2 R^2)\,,\\
{\mathcal R}_3 &=&\frac{17}{12}(\mu_1 R_a^b R_b^c R_c^a+\mu_2 R_{ab}R^{ab}R+\mu_3 R^3)\,.
\end{eqnarray}
Anticipating a relation with the AdS/CFT correspondence, we consider
\be
ds^2=e^{2A(r)}(-dt^2+dx^2)+dr^2\,.
\ee
By demanding that there exists a simple function such that
\be \label{only}
c'(r)=-\frac{T^t_t-T^r_r}{\lp A'^2}\geq 0\,,
\ee
it was shown that $\lambda_2=-3/8 \lambda_1$ and $\mu_1=\frac{64}{17}\mu_3, \mu_2=-\frac{72}{17}\mu_3$. This led to
\be
c(r)=\frac{1}{\lp A'}(1+2\lambda_1 L^2 A'^2+\mu_3 L^4 A'^4)\,,
\ee
satisfying
\be
c'(r)\geq 0\,.
\ee
In the absence of a matter sector
\be \label{norma}
A(r)=\frac{r}{\tilde L}\equiv \frac{r \fin^{1/2}}{L}\,, \qquad 1-\fin+f^2_\infty \lambda_1 +f^3_\infty \mu_3=0\,.
\ee
That AdS is a solution to a higher-derivative theory is not surprising \cite{amsv}. After all, the only effect (if any) of curvature corrections would be to correct the AdS radius. The four derivative theory ($\mu_3=0$) has been studied in \cite{mgads,mgads2,mgads3, Oliva:2009ip}.
Quite curiously, the relative coefficients of the $R^3$ terms work out to be the same as that in the Born-Infeld extension considered in \cite{tekin}. It was further shown in the first paper in \cite{cthms3D} that even the $R^4$ terms work out to be the same in the two approaches. Black hole solutions in the six derivative theory were considered in \cite{nam}. An infinite order generalization of this construction was shown in \cite{paulos}. It was argued in \cite{paulos,me} that by suitably tuning the parameters (in the above example choosing $\lambda_1=-\mu_3 \fin$), the equations of motion for fluctuations work out to be second order. In the six-derivative extension considered above, this implies that $\fin=1$ or in other words the radius of AdS is unaffected. In fact in this case the fluctuations arise from a Fierz-Pauli action of the type \cite{ortin}
\begin{eqnarray}\label{FP}
S&=&\frac{1-\mu_3}{2\lp}\int d^3 x \sqrt{-g}\left(\frac{1}{4} \nabla_\mu h_{\rho\lambda}\nabla^\mu h^{\rho\lambda}-\frac{1}{2}\nabla_\mu h_{\rho\lambda} \nabla^\rho h^{\mu\lambda} +\frac{1}{2} \nabla_\mu h^{\mu\nu} \nabla_\nu h-\frac{1}{4}\nabla_\mu h\nabla^\mu h \right. \nonumber \\ &&~~~~~~~~~~~+ \left. \frac{\Lambda}{2}(h^{\mu\nu}h_{\mu\nu}-\frac{1}{2}h^2)+O(h^3)\right)\,.
\end{eqnarray}
Here $\nabla_\mu$ is with respect to the AdS background metric $g_{\mu\nu}$ with no fluctuations and $h_{\mu\nu}$ is a small fluctuation. Here $h=h_\mu^\mu$. In order to see the equivalence, one needs to choose the transverse traceless gauge.
This is precisely the action one gets when one expands the Einstein-Hilbert action around AdS (or de Sitter) in the presence of a cosmological constant $\Lambda$ (which in our case for AdS is $-1/L^2$). In other words, the combination of the four and six derivative terms are such that the only effect on fluctuations (upto $O(h^2)$) is to renormalize the coupling constant. Since we are in $D=3$ there are no propagating modes and as such there appear to be no restrictions on $(1-\mu_3)$. However the CFT central charge is proportional to $1-\mu_3$ as a result we need $\mu_3<1$ for there to be a sensible AdS/CFT dictionary. Of course, it would have been rather weird if this condition was not satisfied since then the effective Newton constant would be negative! One other interesting point to note about the above construction is that it is crucial to have the cosmological constant to begin with for this to work. An easy way to see this is the following: to get rid of the cosmological constant, we need to send $L\rightarrow \infty$ while rescaling $\mu_3 L^4=\hat \mu, \lambda_1 L^2=\hat\lambda$ keeping $\hat\mu, \hat\lambda$ fixed. Using eq.(\ref{norma}) this would then simply lead to $\fin=0$. This seems to be a peculiar feature of the six derivative extension and is not expected to hold in general \cite{paulos}.

Another point to note in this construction is that if we assume that there is a flow between two asymptotically AdS spaces, then the equations of motion for fluctuations are going to be two derivative {\it only} around one of the spaces. This is because we have tuned the parameters to yield two derivative equations of motion for a specific $\fin$. Once we change $\fin$, which is what will happen in a flow, the equations of motion around the other AdS will no longer be two derivative. This is markedly different from what is possible in $D=4$ and higher where one can tune the parameters to get two derivative fluctuations around any (anti) de Sitter space.

Let us summarize the way the theory was constructed. The motivation above was to have a simple $c$-theorem. Furthermore, we demand that the equations of motion for fluctuations around AdS are two derivative. This is the point of view that we will take to construct lagrangians in $D=4$. If we instead demanded that fluctuations around de Sitter should be two derivative, this is also easily achieved. Either one starts anew by flipping the sign of the cosmological constant and then studying fluctuations around de Sitter, or one replaces $L\rightarrow i L, r\rightarrow i t, t\rightarrow i y$ in the AdS solutions. This leads to the constraint $\lambda_1=\mu_3$ on the parameters as opposed to $\lambda_1=-\mu_3$ in the AdS case. The form for the action for fluctuations is still the same as eq.(\ref{FP}) with $\Lambda=1/L^2$ in this case.

One further nice property of NMG (and the above extension) is the following. It was pointed out in \cite{me} that the $c$-function has a connection with the Wald formula since
\be\label{wald2}
\frac{1}{2}g_{rr}g_{tt}\frac{\partial {\mathcal L}}{\partial R_{rtrt}}=1+2 L^2 \lambda_1 A'(r)^2+L^4\mu_3 A'(r)^4=\lp c(r) A'(r)\,,
\ee
where ${\mathcal L}$ is the lagrangian. In the de Sitter case, there is a similar relation as well. If we write the metric as
\be
ds^2=-dt^2+ e^{\gamma(t) }(dx^2+dy^2)\,,
\ee
then
\be\label{wald3}
\frac{1}{2}g_{xx}g_{tt}\frac{\partial {\mathcal L}}{\partial R_{xtxt}}=1-2 L^2 \lambda_1 H(t)^2+L^4\mu_3 H(t)^4 \equiv \frac{\lp}{2\pi^2} s(t) H(t)\,,
\ee
where $\dot\gamma(t)=H(t)$, the Hubble parameter.
Using the equations of motion and assuming the null energy condition on the additional matter sector, we find that
\be
\dot s(t)\geq 0\,.
\ee
In other words, $s(t)$ is increasing with time. Now in this case $s(t)$ has the natural interpretation of the entropy on the apparent horizon.
To see this recall that Wald's formula for entropy \cite{wald}
\be
S=-2\pi \oint dx \sqrt{h} \frac{\partial {\mathcal L}}{\partial R_{abcd}} \hat \epsilon_{ab} \hat \epsilon_{cd}\,,
\ee
where $\hat \epsilon_{ab}$ is the binormal to the horizon. In this case $S$ works out to be
\be
S=\frac{\pi A}{\lp}g_{xx}g_{tt}\frac{\partial {\mathcal L}}{\partial R_{xtxt}}\,.
\ee
Now the area of the apparent horizon is simply $A=2\pi/H(t)$ since the location of the apparent horizon is at a distance $D=1/H(t)$ and we are in two spatial dimensions. In de Sitter space this gives the correct definition of entropy for Einstein gravity \cite{gibhawk,bousso} on the cosmological horizon. This leads to the interpretation of $s(t)$ as the entropy and we see that if the null-energy condition is satisfied on the additional matter sector (with respect to the full metric) then entropy increases with time.
We would want $s(t)\geq 0$. First consider $\mu_3=0$. This immediately leads to
\be\label{up1}
H^2\leq \frac{1}{2 L^2 \lambda_1}\Rightarrow H L\leq \sqrt{\frac{1}{2 \lambda_1}}\,,
\ee
or in other words, there is an upper bound for $H$ if $\lambda_1 >0$. Turning on $\mu_3=\lambda_1$ this changes to
\be\label{up2}
H L \leq \sqrt{\frac{\lambda_1+\sqrt{\lambda_1(\lambda_1-1)}}{\lambda_1}}\,,\quad {\rm if~~} \lambda_1\geq 1\,.
\ee
One curious feature of this theory is that it is not necessary for area to be increasing for entropy to be increasing. To see this consider for simplicity $\mu_3=0$
\be
\dot s(t)=-\frac{1}{\lp H^2}\dot H(1+2 L^2 \lambda_1 H^2)\geq 0\,.
\ee
Now in standard cosmology with $\lambda_1=\mu_3=0$ this implies $\dot H(t)\leq 0$ or in other words the standard result that the Hubble parameter decreases with time (or $A\propto 1/H(t)$ increases with time). However in the presence of $\lambda_1,\mu_3$ we see that this no longer holds. If $\mu_3=0$ then with $(1+2 L^2 \lambda_1 H(t)^2)<0$ we indeed have $\dot H >0$ or in other words the area decreases with time but due to the higher derivative contribution which overwhelms this decrease, entropy still continues to increase. It will be very interesting to consider this example in the context of cosmological bounce solutions\footnote{The first of eq.(\ref{eoms1}) suggests that for $\rho>0$, in order to have a bounce $\Lambda<0$! It is still possible to have a de Sitter solution in this case, however the fluctuations will now involve more than two derivatives. Bouncing cosmologies in non-relativistic higher derivative theories have been proposed in \cite{bounced}.}. It appears at first sight that this regime will contain ghost excitations.

If we were to consider fluctuations around the standard FRW metric
\be
ds^2=-dt^2+a(t)^2 (dx^2+dy^2)\,,
\ee
then there would be more than two-derivative terms in the equations of motion. All these terms come multiplied by either
$$
(\lambda_1-L^2\mu_3 H^2-2 L^2\mu_3 \dot H)\frac{1}{a^2}\,,
$$
or
$$
(\lambda_1-L^2\mu_3 H^2)\frac{H}{a^2}\,.
$$
Here  $H(t)$ satisfies
\be\label{eoms1}
H^2-\Lambda+L^2\lambda_1 H^4-L^4\mu_3 H^6=8\pi \rho\,,\quad \dot H(1+2 L^2 \lambda_1 H^2-3 L^4\mu_3 H^4)=-8\pi (\rho+P)\,.
\ee
Thus in the absence of matter ($\rho=P=0$) if $H=\sqrt{\Lambda}=1/L$ with $\lambda_1=\mu_3$, then the potentially problematic three and higher derivative terms would disappear from the equations of motion and we would be left with the promised two derivative equations for the fluctuations. Furthermore, when we add matter we would expect $a(t) \sim t^p$ so that $H=p/t$ in which case the ghost terms would be proportional to either $(\lambda_1-L^2\mu_3 p^2/t^2)p/t^{2p+1}$ or $(\lambda_1+L^2\mu_3 p^2/t^2)/t^{2p}$. In both cases as long as $p>0$ these terms would become irrelevant at large times. Also note that as in the AdS case, the fluctuations are two derivative only around one of the de Sitter vacua if there is a flow between two de Sitter vacua.

Let me end this section by pointing out that Schwarzschild de Sitter solutions are straightforward to find. It is easy to check that
\be
ds^2=-f(r)dt^2+\frac{dr^2}{f(r)}+r^2 d\phi^2\,,
\ee
with $f(r)=1-2m -r^2 H_0^2$ satisfy the equations of motion where $H_0^2-\Lambda+L^2\lambda_1 H_0^4-L^4\mu_3 H_0^6=0$. In de Sitter in general dimensions, there is a maximum mass that the black hole can have which is fixed by the fact that in this case the de Sitter horizon and black hole horizon coincide. In $D=3$ there is only the de Sitter horizon \cite{dscft}. When $f(r_h)=0$ for $r_h=\sqrt{1-2m}/H_0$, there is a conical singularity with a positive deficit angle at the origin corresponding to a point-like mass at the south pole. Further there is a maximum value of $m=1/2$.  The Wald entropy is given by
\be
S_{Wald}=\frac{2\pi A}{\lp} (1-2 L^2 \lambda_1 H_0^2+L^4\mu_3 H_0^4)\,,
\ee
with $A=2\pi r_h=2\pi \sqrt{1-2m}/H_0$
which further justifies the identification of $s(t)$ in eq.(\ref{wald3}) as the entropy of de Sitter space.

\vskip 1cm
\noindent \underline{{\bf Comment on the sign of $\lambda_1$}}\\
Let me briefly comment on the sign of $\lambda_1$ when $\mu_3=0$. There are two interesting possibilities for either sign of $\lambda_1$. If $\lambda_1>0$, then $(1+2 L^2 \lambda_1 H^2)>0$ and we have the sign of the kinetic term to be positive and an upper bound on the Hubble parameter set by eq.(\ref{up1}) arising from demanding positive entropy. If $\lambda_1<0$ then $(1+2 L^2 \lambda_1 H^2)$ will run the risk of turning negative as $H$ increases and introducing ghosts. In this case there will be no upper bound on the Hubble parameter although entropy is positive. Furthermore combined with eq.(\ref{up1}) or eq.(\ref{up2}) and the first of eq.(\ref{eoms1}) it is easy to see that $a(t)$ cannot be zero for radiation or dust matter. In this case $\rho\sim 1/a^2$ or $\rho \sim 1/a^3$ which blows up. Thus to be consistent $H\rightarrow \infty$ but this is forbidden by an upper bound on $H(t)$. As such aesthetically it may seem that $\lambda_1>0$ is a more pleasing possibility! It will be interesting to understand what goes wrong when the upper bound on $H(t)$ does not hold and $s(t)$ becomes negative.

\section{Higher derivative gravity in $D=4$}
It is interesting to ask the following: Given an action for the fluctuations around (anti) de-Sitter of the form eq.(\ref{FP}), what is the covariant action whose quadratic expansion, $O(h^2)$ would give rise to this? If we worked using a truncated lagrangian which included upto only four derivative terms, the answer is that only Einstein-Hilbert with a cosmological constant would work. This is because $R^2$ would generally lead to four derivative equations of motion. In $D=3$ this is not necessarily problematic: the new massive gravity is consistent under certain circumstances. However, in higher dimensions only Gauss-Bonnet or Lovelock terms would lead to ghost-free theories. In $D=4$ this is not interesting since Gauss-Bonnet would not contribute to the equations of motion (however, the Gauss-Bonnet terms can contribute to the entropy in a crucial way as we will discuss in the next section). Although $f(GB)$ theories have been considered in the literature (see \cite{fofR} for references), these theories are not viable for cosmology. This gives us motivation for searching for other creative theories.



In $D=4$, if one expanded around flat space, the equations of motion for a generic four derivative gravity theory will always be more than two derivatives\footnote{For $R+R^2$, the theory can be mapped to Einstein gravity and scalar field which is free of the usual problems \cite{fofR}.}. It is in fact easy to argue that adding even higher curvature terms would be of little use since the equations of motion would typically involve $h R^2\sim 0$. However, around a background like (anti) de Sitter, the story is not so simple. In fact as we saw in the $D=3$ example, one can play the $R^2$ terms off the $R^3$ terms to have two derivative equations of motion. We are now going to show that there is a similar construction in $D=4$. One nice feature of this theory, unlike the $D=3$ example, is that one can have two derivative fluctuations around any (anti) de Sitter space.

Our starting point is the action given in eq.(\ref{action}) in $D=4$.
Although we have left an explicit cosmological constant in the lagrangian, we will soon argue that this is not necessary to have a (anti) de Sitter solution with two derivative fluctuations unlike the $D=3$ case. Let us begin with the AdS case. The metric is
\be\label{ans1}
ds^2=e^{2 A(r)}(-dt^2+ dx^2+dy^2)+dr^2\,,
\ee
where $A(r)=r \sqrt{\fin}/L$.
The parameters need to satisfy the following constraints in order for the equations of motion for the fluctuations to be two derivative\footnote{In a previous version of this paper it was erroneously implied that these constraints are necessary for two derivative equations of motion for the fluctuations. While these are {\it sufficient} conditions, these are not {\it necessary}. In fact it is shown in \cite{future1}, that it is possible to have two derivative equations of motion {\it without} having a simple $c$-theorem.} and such that a simple $c$-theorem exists:
\beqa \label{constr3}
\la_1+\la_2+3\la_3&=&0 \,,\\
3\mu_1+48\mu_2+14\mu_3+16 \mu_4+18\mu_5+60\mu_6+216\mu_7&=&0 \,,\nonumber \\
12\mu_2+4 \mu_3+3 \mu_4+5 \mu_5+12 \mu_6 +36 \mu_7 &=& 0\,, \nonumber \\
\lambda_2+4\lambda_1-3\fin(-8\mu_2-2\mu_3-\mu_4+\mu_5) &=&0\nonumber \,.
\eeqa
Note that using eq.(\ref{idX}) we could set $\mu_7=0$.
The first three constraints will be imposed explicitly while we will leave the last one implicit in what follows. By imposing all the constraints we will find theories that have second order linearized equations of motion and admit a simple $c$-theorem. The first three constraints were obtained by demanding that a simple $c$-theorem exists as in the $D=3$ case. The last constraint is needed to make the linearized equations of motion two derivative. This will also serve the purpose of illustrating that it is {\it not necessary} to demand two derivative equations of motion for fluctuations to have $c$-theorems. The first of the constraints is satisfied both by Gauss-Bonnet and by Weyl-squared. In the absence of the six-derivative terms, the last constraint would have uniquely selected Gauss-Bonnet as one would have naively guessed. However, the interesting thing to note here is that in the presence of six-derivative terms,
it is not necessary for the $R^2$ terms to be Gauss-Bonnet to have two derivative fluctuations. Of course, this is a highly contrived situation where the couplings $\lambda_i, \mu_i$'s have to be fine-tuned to make things work. If the $R^2$ and $R^3$ couplings were independent of one another then we would be forced to choose the Gauss-Bonnet term. Furthermore, using Gauss-Bonnet and $W_1$ as defined in eq.(\ref{Winv}) will satisfy all the constraints as well. If we were to use $W_1$ however, $\fin=-\Lambda L^2/3$ and there is no way to get AdS without an explicit cosmological constant in the first place. We would like to keep things as general as possible, not only because of this, but also to explore the possibility of having exact black hole solutions. Another point to note is that with $W_1$ the $c$-function would not receive corrections from the six derivative terms. So we will keep things more general for now.
Here in the absence of matter $\fin$ satisfies
\be\label{fineq}
1-\fin -(\mu_1+\mu_4-\mu_5)f_\infty^3=0\,.
\ee
Thus we have a five-parameter family of theories. Using the equations of motion and the null energy condition one can show that $c(r)$ defined through \cite{future1}
 \be
 \frac{1}{2}g_{rr}g_{tt}\frac{\partial {\mathcal L}}{\partial R_{rtrt}}=[1+2(2\lambda_1+\lambda_2)L^2 A'^2-3(\mu_1+\mu_4-\mu_5)L^4 A'^4)]=\lp^2 A'^2 c(r)\,,
 \ee
 is monotonically increasing. In other words $c_{UV}>c_{IR}$ when there are UV and IR fixed points. Another point to note is that as in \cite{mr}, one can infer unitarity from the left hand side of eq.(\ref{fineq}). The action for the fluctuations take the form $(1+3(\mu_1+\mu_4-\mu_5)\fin^2)$ times what we have in Einstein gravity. As a result,  $(1+3(\mu_1+\mu_4-\mu_5)\fin^2)>0$ needs to hold for unitarity. Together with eq.(\ref{fineq}) this translates into the condition $\mu_1+\mu_4-\mu_5>-4/27$.

 As in the $D=3$ case one will have two derivative fluctuations only around one of the AdS vacua if there is a flow between two different AdS vacua.  However unlike the $D=3$ case, we have a bit more freedom here. In fact in addition to eq.(\ref{constr3}) if $\mu_5-\mu_4=8\mu_2+2\mu_3$, then the fluctuations around any AdS vacua will be two derivative! In this case, the last constraint in eq.(\ref{constr3}) gives us the familiar result that the $R^2$ combination will be Gauss-Bonnet (and hence topological).

 In fact we can do much better. It is possible to actually have the fluctuations around any $A(r)$ in eq.(\ref{ans1}) to be two derivative and in fact the same as in Einstein gravity. In a sense this is exactly like Gauss-Bonnet but at six derivative order! The choice of parameters that accomplishes this is
 \be\label{2derany}
\lambda_2=-4 \lambda_1=-4\lambda_3\,,\quad \mu_7=\frac{\mu_1}{36}+\mu_2, \quad \mu_5=8 \mu_2=-\mu_6,\quad \mu_4=-\mu_1+8\mu_2=-2\mu_3\,.
  \ee
  If we use eq.(\ref{idX}) and set $\mu_7=0$ then
   we will have a two parameter family of lagrangians with nice properties. Now, unlike Gauss-Bonnet which is topological and does not enter any equation of motion in $D=4$, the six derivative terms are more interesting. Only under further conditions does one find simple exact black hole solutions as we will mention shortly. In other words, the six derivative terms are not topological. With this choice of parameters in empty AdS, $\fin=1$. This finding will continue to hold for fluctuations around de Sitter as well. Also note that $\alpha W_1$ satisfies the constraints in eq.(\ref{2derany}). However, in what follows, unless explicitly specified, we will only impose the first three constraints in eq.(\ref{constr3}).

 In order to see that it is not necessary to have the explicit cosmological constant in the theory, we rescale $\hat\lambda_i=\lambda_i L^2, \hat \mu_i=\mu_i L^4$ and send $L\rightarrow \infty$. Now the AdS radius works out to be $\fin/L^2={\hat f}_\infty$. In this case, ${\hat f}_\infty$ satisfies
\be {\hat f}_\infty+(\hat \mu_1+\hat \mu_4-\hat \mu_5){\hat f}_\infty^3=0\,. \ee Now it is possible to have
${\hat f}_\infty = 1/\sqrt{\hat \mu_5-\hat \mu_4-\hat \mu_1}$ as a non-trivial solution. However, as in
Boulware and Deser \cite{bd} fluctuations around these vacua will contain ghosts. Note that
$\fin$ does not depend on the $R^2$ parameters although the fluctuations do.

It is also possible to get exact black hole solutions\footnote{These solutions would not have been found using the method in \cite{mr} since the equations of motion involving $f(r)$ in our case is fourth order but an exact solution nonetheless exists. The method in \cite{mr} will only work when the $f(r)$ has second order equations of motion.} if in addition to the first three constraints in eq.(\ref{constr3}) we restrict $\mu_1=0,4\mu_2+\mu_3=0$. In this case it is possible to show that
\be
ds^2=-f(r)dt^2+\frac{dr^2}{f(r)}+\frac{r^2}{L^2}(dx^2+dy^2)\,,
\ee
with $f(r)=\fin r^2/L^2 (1-M/r^3)$, with $\fin$ obeying eq.(\ref{fineq}), satisfies the equations of motion. The equations of motion involving $f(r)$ are four derivative. In spite of this we have a remarkably simple solution! In fact apart from the corrected asymptotic AdS radius, the solution is of the same form as in Einstein gravity. We do not yet have an understanding for this simplicity. The complicated nature of the equations of motion also makes the proof of Birkhoff's theorem along the lines of \cite{Oliva:2010zd} a difficult problem. We have a three parameter family of exact solutions. It is not possible to use Weyl invariants to accomplish this since setting $\mu_1=0$ would need $\alpha=-\beta$ in $\alpha W_1+\beta W_2$ which vanishes as it is a Schouten identity. We have here more general exact black hole solutions where $\fin\neq 1$ (if we use the last constraint in eq.(\ref{constr3}) then $\fin=1$). The ratio of shear viscosity to entropy density $\eta/s$ in this model can be calculated following for example \cite{mps}. It works out to be $1/4\pi$ in spite of the fact that there are higher derivative terms. $\eta$ and $s$ each get corrected but in the same manner. It will be interesting to constrain the parameter space following \cite{mps} and figure out which hydrodynamic quantities do get corrected. We will leave the possibility of exact black holes in theories with $\mu_1\neq 0, \mu_3\neq -4\mu_2$ as an interesting open question.

We can easily extend the analysis to de Sitter space. We find that
\be
ds^2=e^{2 H_0 t}(dx^2+dy^2+dz^2)-dt^2\,,
\ee
is a solution with $H_0$ satisfying
\be\label{H0cond}
\frac{\Lambda L^2}{3}-H_0^2 L^2 -(\mu_1+\mu_4-\mu_5) H_0^6 L^6=0\,.
\ee
 To get two derivative equations of motion for the fluctuations, the parameters satisfy the following constraints:
\beqa \label{constr4}
\la_1+\la_2+3\la_3&=&0 \,,\\
3\mu_1+48\mu_2+14\mu_3+16 \mu_4+18\mu_5+60\mu_6+216\mu_7&=&0 \,,\nonumber\\
12\mu_2+4 \mu_3+3 \mu_4+5 \mu_5+12 \mu_6 +36 \mu_7 &=& 0\,, \nonumber \\
\lambda_2+4\lambda_1+3 H_0^2 L^2(-8\mu_2-2\mu_3-\mu_4+\mu_5) &=&0 \nonumber\,. \eeqa We will impose the first
three of these constraints explicitly and leave the last one implicit. Furthermore, if
$\mu_5-\mu_4=8\mu_2+2\mu_3$, then the fluctuations around any de Sitter will be two derivative while if
eq.(\ref{2derany}) holds then the fluctuations around any FRW will be two derivative. As before we will not
impose these constraints unless explicitly specified. Again we can rescale the original cosmological parameter
out by taking $L\rightarrow \infty$. Now $H_0=1/\sqrt{\hat \mu_5-\hat \mu_4-\hat \mu_1}$ although the resulting
vacua will again contain ghosts. One can also construct exact asymptotically de Sitter Schwarzschild solutions
if $\mu_1=0,4\mu_2+\mu_3=0$. In this case it is possible to show that \be\label{sds}
ds^2=-f(r)dt^2+\frac{dr^2}{f(r)}+r^2 (d\theta^2+\sin^2\theta d\phi^2)\,, \ee is an exact solution with
$f(r)=(1-2m/r -r^2 H_0^2)$ with $H_0$ satisfying eq.(\ref{H0cond}). The asymptotic space, $r\rightarrow \infty$
is de Sitter in static coordinates \cite{dscft}. The Wald entropy\footnote{The fact that Gauss-Bonnet
contributes to Wald entropy in 4 dimensions was pointed out in a different context in \cite{krishnan}.} of
these black holes works out to be
\begin{eqnarray}\label{waldent}
S_{Wald}&=& \frac{2 \pi A}{\lp^2}\left[-\frac{4 L^2 m \lambda_1}{r_h^3}+\left\{1-2(2\lambda_1+\lambda_2)H_0^2 L^2-3 (\mu_4-\mu_5)H_0^4 L^4\right\}\right]\,,\\
&\equiv & \frac{2 \pi A}{\lp^2}\left[-\frac{4 L^2 m \lambda_1}{r_h^3}+s_0 \lp^3 H_0^2\right]\,,
\end{eqnarray}
where $A=4\pi r_h^2$ is the area of the horizon and $r_h$ is the location of the black hole horizon. The second equation defines $s_0$ whose meaning will be clear in the next section. Now de Sitter space has an interesting property. There is a maximum mass black hole that one can fit inside de Sitter \cite{dscft}. This happens when the de Sitter horizon and the black hole horizon coincide. The location of the horizon is found by solving the depressed cubic equation $f(r)=0$. When $m=1/(3 \sqrt{3} H_0)$, then there are two equal positive roots $r_h=1/(\sqrt{3} H_0)$, in other words the de Sitter horizon and the black hole horizon are on top of each other\footnote{$m/r_h^3$ is maximized for this choice as well.}. Thus demanding that the black hole entropy is positive we get the following interesting inequality
\be\label{entbound}
s_0\geq \frac{4 L^2}{\lp^2}\lambda_1\,.
\ee
In the next section we will see that $s_0$ is the entropy of de Sitter space. Thus we seem to find that there is some minimum entropy in the system if $\lambda_1>0$! In $D=3$, there are exact Schwarzschild de Sitter solutions as well (with no constraints on $\lambda_1$ or $\mu_3$) but their Wald entropy does not receive any contribution analogous to the $\lambda_1 m$ term in $D=4$. Hence eq.(\ref{entbound}) seems special to $D=4$.

\section{Entropy theorems in cosmology}
Since $c$-theorems in the context of AdS/CFT will be discussed extensively in \cite{future1}, we will focus on $c$-theorems in the context of cosmology. But before we begin, let me briefly summarize the findings in \cite{future1}. In \cite{ms}, we found that in the context of the quasitopological gravity introduced in \cite{mr}, there are nice $c$-theorems in arbitrary dimensions. In odd dimensional CFTs ($D=4$ for e.g.) the quantity that is flowing was interpreted as entanglement entropy of the CFT on $S^{D-2}\times R$. In \cite{future1}, it is shown that these $c$-theorems can be extended for a wide class of higher derivative gravity theories. In the context of cosmology, we naturally interpret the quantity which flows as the entanglement entropy in de Sitter space across the cosmological horizon. Let us consider this calculation using the static coordinates since in these coordinates, the results can be compared directly with those presented in \cite{ms} in the AdS/CFT context. The calculation can be done using the metric eq.(\ref{sds}) which leads to eq.(\ref{waldent}). The de Sitter entropy is extracted by setting $m=0$ in eq.(\ref{waldent}). The interpretation of the AdS/CFT calculation was that the $c$-function at the fixed points gives information about the entanglement entropy between the two halves of $S^{2}$ when the CFT is placed on $S^{2}\times R$. In the de Sitter context, the interpretation\footnote{In the dS/CFT context, it is again mapped onto the entanglement entropy of the two halves of $S^{2}$.} on the bulk side is straightforward. It is simply the entanglement entropy of the two disconnected regions\footnote{The horizons are at $r=1/H_0$. As discussed in \cite{dscft}, no single observer can access the entire de Sitter spacetime and the entropy in this context is the entropy of whatever is behind the horizon.
} of de Sitter space.

Now let us turn to $c$-theorems in cosmology. For this we want to start with a metric of the form
\be
ds^2=-dt^2+e^{\gamma(t)}(dx^2+dy^2+dz^2)\,.
\ee
Now we calculate
\be
\frac{1}{2}g_{xx}g_{tt}\frac{\partial {\mathcal L}}{\partial R_{xtxt}}=1-2(2\lambda_1+\la_2)L^2 H(t)^2-3 (\mu_1+\mu_4-\mu_5)L^4 H(t)^4=\frac{\lp^2}{4\pi^2} s(t) H(t)^2\,,
\ee
where $s(t)$ is a monotonic function which follows from a straightforward application of the equations of motion and assuming the null energy condition (assuming $H(t)>0$). We see that this is in agreement with eq.(\ref{waldent}) in the zero mass limit as expected. One important thing to note here is the following. Let us impose eq.(\ref{2derany}) so that the six derivative terms do not contribute to the entropy and the $R^2$ combination is just the Gauss-Bonnet term. The equations of motion do not get affected due to this term. However, the entropy does. The entropy here is given by
\be\label{entGB}
s(t)|_{GB}=\frac{4\pi^2}{\lp^2 H(t)^2}(1+4\lambda_1 L^2 H(t)^2)\,.
\ee
As in the new massive gravity example, we have an upper bound for $H\leq \sqrt{-1/(4\lambda_1)}$ if $\lambda_1<0$. It will be interesting to ask what goes wrong when the entropy becomes negative. The analysis in \cite{parikh} may have some relevance in this case. Note that eq.(\ref{entGB}) also implies $\displaystyle\frac{s(t)|_{GB}}{4\pi^2}\equiv s_0\geq 4 L^2 \lambda_1/\lp^2$ which is exactly what we found in eq.(\ref{entbound}). Entropy of de Sitter spaces have been investigated previously in four derivative theories in general dimensions in \cite{shu} following \cite{cai} and our conclusion is in agreement with their analysis.

\section{Discussion}
We considered higher derivative lagrangians in $D=3$ and $D=4$ which were constructed such that fluctuations around (anti) de Sitter had two derivative equations of motion and such that these theories admit simple $c$-theorems both in the context of AdS/CFT and cosmology along the lines of \cite{me,ms,future1}. The most general such lagrangian in $D=4$ [see eq.(\ref{action}) subject to eq.(\ref{constr3}) or eq.(\ref{constr4})] had five parameters. In addition if we imposed $\mu_5-\mu_4=8\mu_2+2\mu_3$, then the fluctuations around any (anti) de Sitter space have two derivative equations of motion. Exact black hole solutions\footnote{Actually exact black hole solutions of the type discussed here exist even if eq.(\ref{constr3}) or eq.(\ref{constr4}) were not imposed and only $\mu_1=0,4\mu_2+\mu_3=0$ were imposed.} were found if $\mu_1=0,4\mu_2+\mu_3=0$ along with eq.(\ref{constr3}) or eq.(\ref{constr4}). Unlike \cite{mr} these black holes were similar what to what is found in Einstein gravity except that the asymptotic (anti) de Sitter radius gets corrected. It will be interesting to analyze these theories in detail as in \cite{mps}. As was noted in \cite{mr}, the construction outlined there does not extend to $D=4$ or $D=6$. The approach used in this paper following \cite{future1} is a useful way to extend their theories to these dimensions. In the course of our work, we also found that if we asked for two derivative equations of motion for fluctuations around any FRW background or around a static domain wall in $D=4$, then the most general lagrangian upto six derivative curvature invariants is $L=\sqrt{-g}(R-2\Lambda+ \lambda GB +\alpha W_1)$ where $W_1$ is defined in eq.(\ref{Winv}). There is no analogous solution\footnote{In $D=3$ the role of the Weyl tensor is played by the Cotton tensor $C^{ab}=\epsilon^{a c d}\nabla_c(R^b_d-\frac{1}{4}\delta^b_d R)$. Had we considered theories made of invariants of the Cotton tensor alone, then it is possible that the fluctuations around a general FRW would be two derivative. However, there is no way of making the $R^2$ contributions to the fluctuations cancel on their own.} in $D=3$.

In $D=4$ an inequality was found which suggests that there is a lower bound for the de Sitter entropy. This arises due to the contribution of the $R_{abcd}R^{abcd}$ term in the lagrangian and is present even though the equations of motion for the background
do not receive contributions from this term. This is a curious feature of $D=4$ and it will be interesting to investigate an analogous feature in $D=6$ due to $R^3$ terms which also will not contribute to the equations of motion but will contribute to the entropy via Wald's formula. Of course fluctuations around the background may be sensitive to these terms in general except when the combination is Gauss-Bonnet in $D=4$ or Lovelock in $D=6$. The lower bound is sensitive to the sign of the $R_{abcd}R^{abcd}$ term and exists only if the coefficient is positive. In string theory, scattering amplitude calculations in heterotic or higher curvature corrections to the D-brane effective action both lead to positive coefficients \cite{gwitt,zwie,posr2}. This makes a closer analysis of this bound very interesting. At the level of this paper, it is not clear how general this bound is or if it is specific to the set of lagrangians permitting exact black holes considered in here\footnote{In the context of AdS/CFT, the effect of topological terms has been investigated in \cite{olea1,olea2}. In 4 dimensions, it was found that the Gauss-Bonnet coupling is fixed to a specific (positive) value. The same finding holds for Lovelock terms in higher dimensions.}.

It will also be interesting to carry out a more systematic study of perturbations using these lagrangians. If eq.(\ref{2derany}) is not imposed, perturbations around FRW have fourth order equations of motion, in general we expect there to be ghost modes for these lagrangians. However, as we explicitly showed for $D=3$ the coefficients of these higher order terms become more and more irrelevant as the universe expands. As such, there is a possibility that these many of these models may give sensible cosmology.

We are taking a bottom-up viewpoint \cite{mr, mps} for the construction of these higher derivative lagrangians as there is very little to go by from fundamental theories like string theory. The rules that sensible lagrangians must obey are not clear. For instance, imposing the null energy condition is an assumption which need not be true. However, seeing that the entropy of de Sitter satisfies a monotonicity property when this holds, makes it worthwhile to think that this is a feature that sensible lagrangians must obey. Furthermore, demanding that the fluctuations satisfy two derivative equations of motion is certainly not a prerequisite for the $c$-theorems. In fact these theorems hold even though the last equations of (\ref{constr3}) and (\ref{constr4}) are not obeyed. As such it seems like a worthwhile pursuit to investigate these models further with the hope to learn how to constrain them.

\section*{Acknowledgments}
 I thank Dileep Jatkar, Rob Myers and Miguel Paulos for useful discussions and comments on the draft. I also thank Sumit Das, Claudia de Rham and Louis Leblond for useful discussions and to Bayram Tekin for useful correpsondence. I am especially grateful to Rob Myers for suggesting that sensible lagrangians can be constructed using the approach in this paper and for critical comments on the manuscript.
 Research at Perimeter Institute is supported by the Government of Canada through Industry Canada and by the Province of
Ontario through the Ministry of Research \& Innovation.

\end{document}